\documentclass[aps,prl,twocolumn,showpacs,superscriptaddress,groupedaddress]{revtex4}  
\usepackage{graphicx}  
\usepackage{dcolumn}   
\usepackage{bm}        
\usepackage{amssymb}   
\usepackage{color}

\begin{document}

\widetext

\title{Thermal convection in huddling emperor penguins}

\author{Dmitry Bratsun}
\author{Kirill Kostarev} 
\affiliation{Department of Applied Physics,
Perm National Research Polytechnical University, 614990, Perm, Russia}


\begin{abstract}
Emperor penguins are the only penguin species that winter in Antarctica. As is known, during cold weather, birds huddle together to share body heat. We developed a microscopic model in which penguins interact with each other through an effective potential that describes the birds' intention to move along the gradient of the thermal field. The model describes the aggregation of penguins into a motionless huddle, as observed previously. More interestingly, we found that increasing the number of birds leads to a second-order phase transition, characterized by the excitation of a vortex motion in a huddle. The dynamic behavior ensures a more efficient redistribution of heat between penguins and, consequently, the survival of all birds in the flock. To study the instability mechanism, we developed a continuous model and applied both linear and weakly nonlinear analysis. Numerically, we studied the increasing complexity of vortex structures in a fluidized huddle with increasing its power. Finally, we demonstrate that the effect of sudden fluidization of a huddle, resulting in the spontaneous motion of penguins from the edge of the crowd to its center and back, is essentially thermal convection. The findings are juxtaposed against observations of emperor penguins.
\end{abstract}


\maketitle

Thermal convection, the ordered motion of molecules of a non-uniformly heated liquid or gas induced by the buoyancy, occupies a special place in the history of physics. The study of this phenomenon led to the development of the theory of convective stability~\cite{Rayleigh1916}, the first observations of spatial patterns~\cite{Benard1900}, and the discovery of deterministic chaos~\cite{Lorenz1963}. The concept was used to explain the formation of the Earth~\cite{Nguyen2013}, ocean currents~\cite{Holmes1931}, stellar astrophysics~\cite{Behrend2001}, and RDC processes~\cite{Bratsun2021}, among others.

The standard explanation of natural convection onset within fluid mechanics is as follows: if a small parcel of a stratified fluid is accidentally displaced to a colder environment, it continues to move further away from its equilibrium position due to the Archimedes force~\cite{Gershuni1976}, indicating that the basic stratification is unstable. 
Thus, convective instability is driven by the change in density of the fluid as it heats up and cools down. Prigogine provided a deeper interpretation of the phenomenon based on the synergetics ideas~\cite{Prigogine1984}. Here, one observes a second-order phase transition, enabling the system to adopt a more efficient spatial configuration in response to an increase in the heat flux. But what would change if we go beyond physics and replace molecules with more complex elements? The phenomenon of bioconvection, the pattern-forming motions that occur in suspensions of swimming microorganisms, partially answers this question~\cite{Hill2005}. The capacity of microorganisms to move autonomously during chemotaxis creates a force of buoyancy, causing an instability. Thus, bioconvection is not fundamentally different from thermal convection, though it occurs in an isothermal medium. From a biological perspective, the onset of spatial circulations of bacteria allows them to access a resource essential for the survival of the entire colony.
In the case of higher animals, researchers find it troublesome to distinguish the circulations of animals from their social behavior. As is known, collective behavior can serve as a mechanism for energy conservation, playing a crucial role in the survival of some species~\cite{Trenchard2016}.
However, a simple explanation based on physical mechanisms is no longer sufficient since the primary trigger for such behavior is social adaptation rather than an unconditioned reflex~\cite{Sumpter2010}.

Nevertheless, we can point to one important exception, where higher animals demonstrate collective behavior that has even more similarities with the classical thermal convection than the bioconvection of microbes. It is the huddling behavior of emperor penguins ({\it Aptenodytes forsteri}) during wintering in Antarctica~\cite{Williams1995}.
Until recently, scientists had limited knowledge about the lifecycle of this species due to the difficulty of accessing its habitat.
In the last century, the results of early observations during rare expeditions were summarized in~\cite{Maho1977}, mentioning the collective strategy of these birds for survival. In 2005, the documentary film~\cite{Marche2005} was released, stimulating further research. Among other things, the film demonstrated the spontaneous huddling of penguins when the ambient temperature drops sharply.

In recent years, significant results have emerged from field observations, which also include quantitative measurements~\cite{Ancel1997, Ancel2009, Ancel2015, Richter2018a, Richter2018b, Fretwell2021}.
Air temperature and wind speed are two critical factors that influence the onset of huddling~\cite{Richter2018b}. 
One also suggested that there is a similarity between the huddling and a first-order phase transition, leading to a change in the aggregation state of the medium. Several papers have developed mathematical models to describe the behavior of penguins, with a particular focus on the huddling process~\cite{Waters2012, Gerum2013, Dhiman2018, Gu2018, Harris2023}. More significant second-order phase transitions associated with self-organization have received insufficient attention from researchers. Meanwhile, in several episodes of~\cite{Marche2005} and supplementary video materials~\cite{Richter2018b}, one could observe the impact of sudden fluidization of a dense huddle and the spontaneous excitation of global circulations of penguins within the flock. Biologists refer to this phenomenon as ``social thermoregulation", but, in our opinion, as in the case of bioconvection, the instability mechanism can be explained by purely physical reasons.

In this Communication, we state for the first time that the spontaneous excitation of penguins' circulations in a fluidized huddle is nothing other than a second-order phase transition in the form of the excitation of thermal convection. We develop mathematical models that provide both a microscopic and continuous description of the phenomenon observed in emperor penguins.

\begin{figure*}
\includegraphics[scale=0.09] {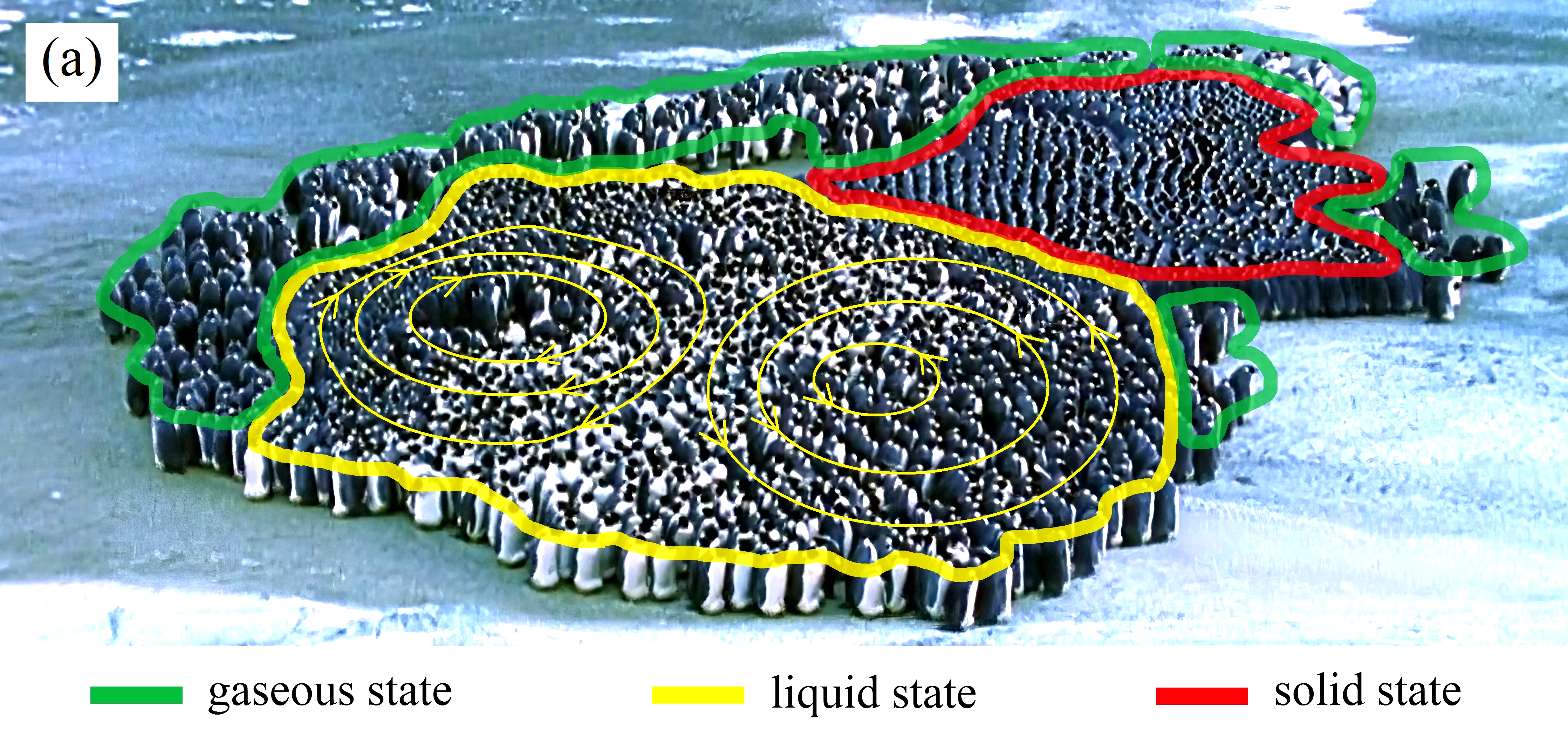}\quad
\includegraphics[scale=0.07]{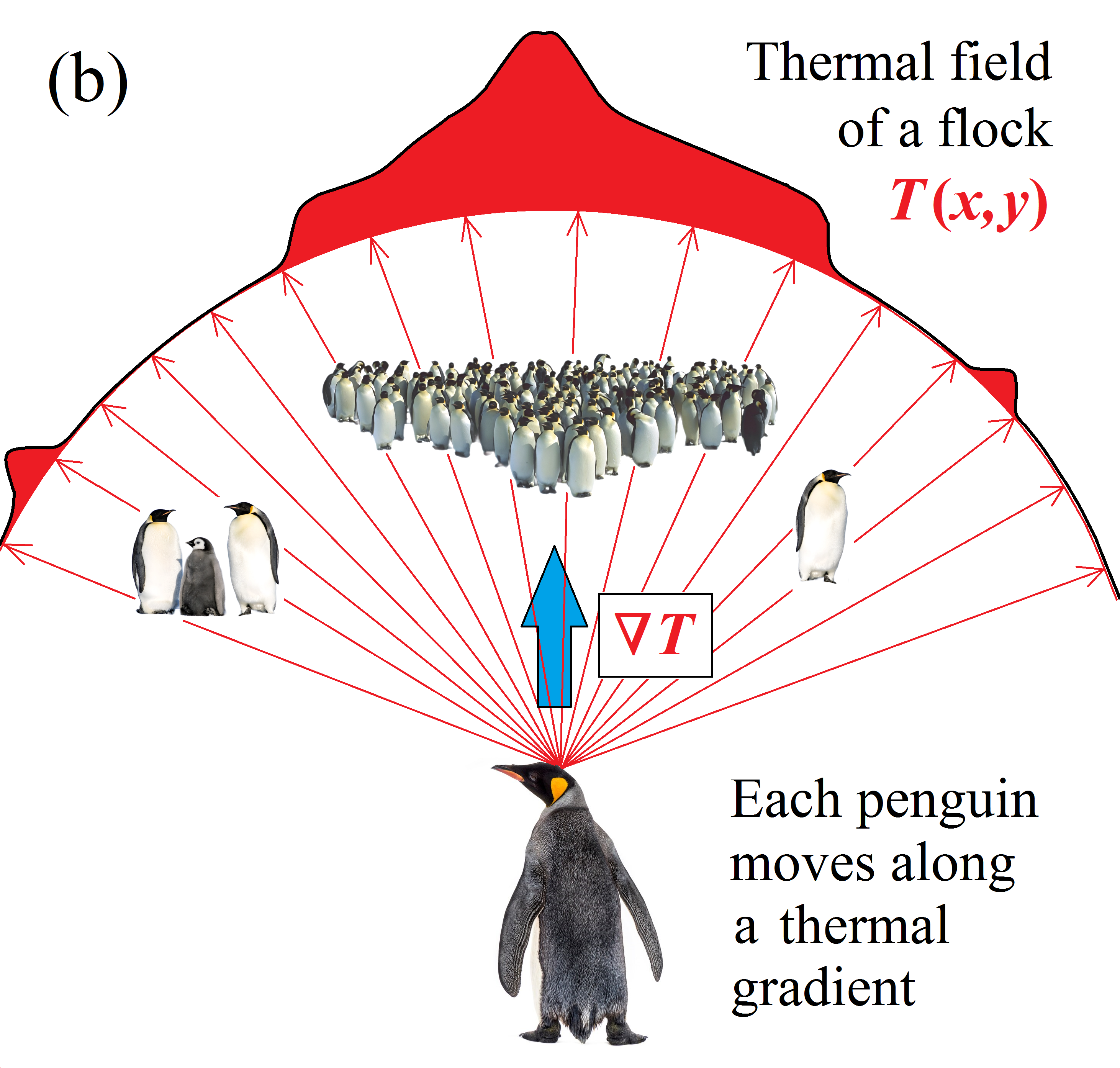}\\
\vspace{1mm}\includegraphics[scale=0.16] {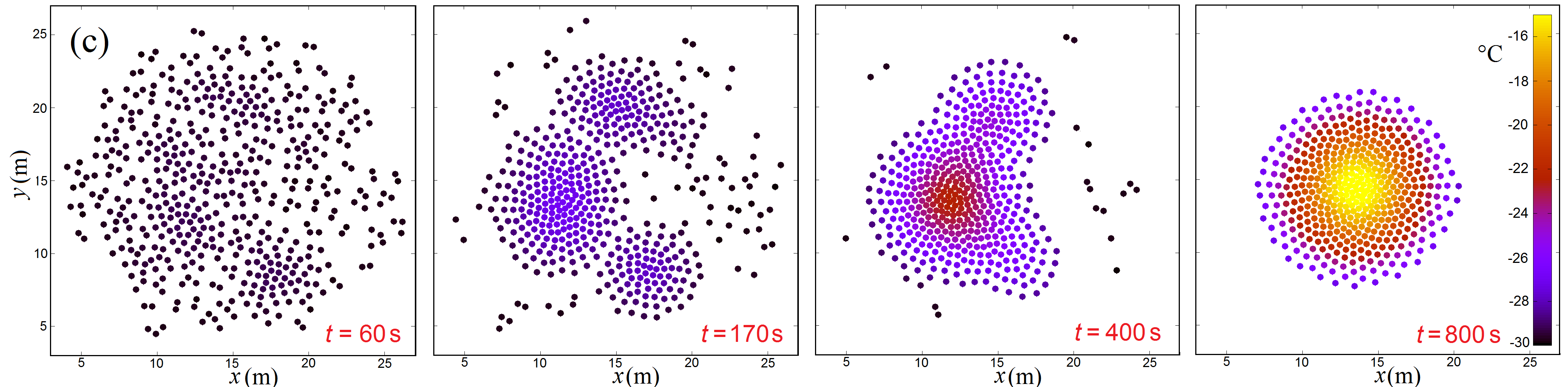}\\
\vspace{1mm}\includegraphics[scale=0.12] {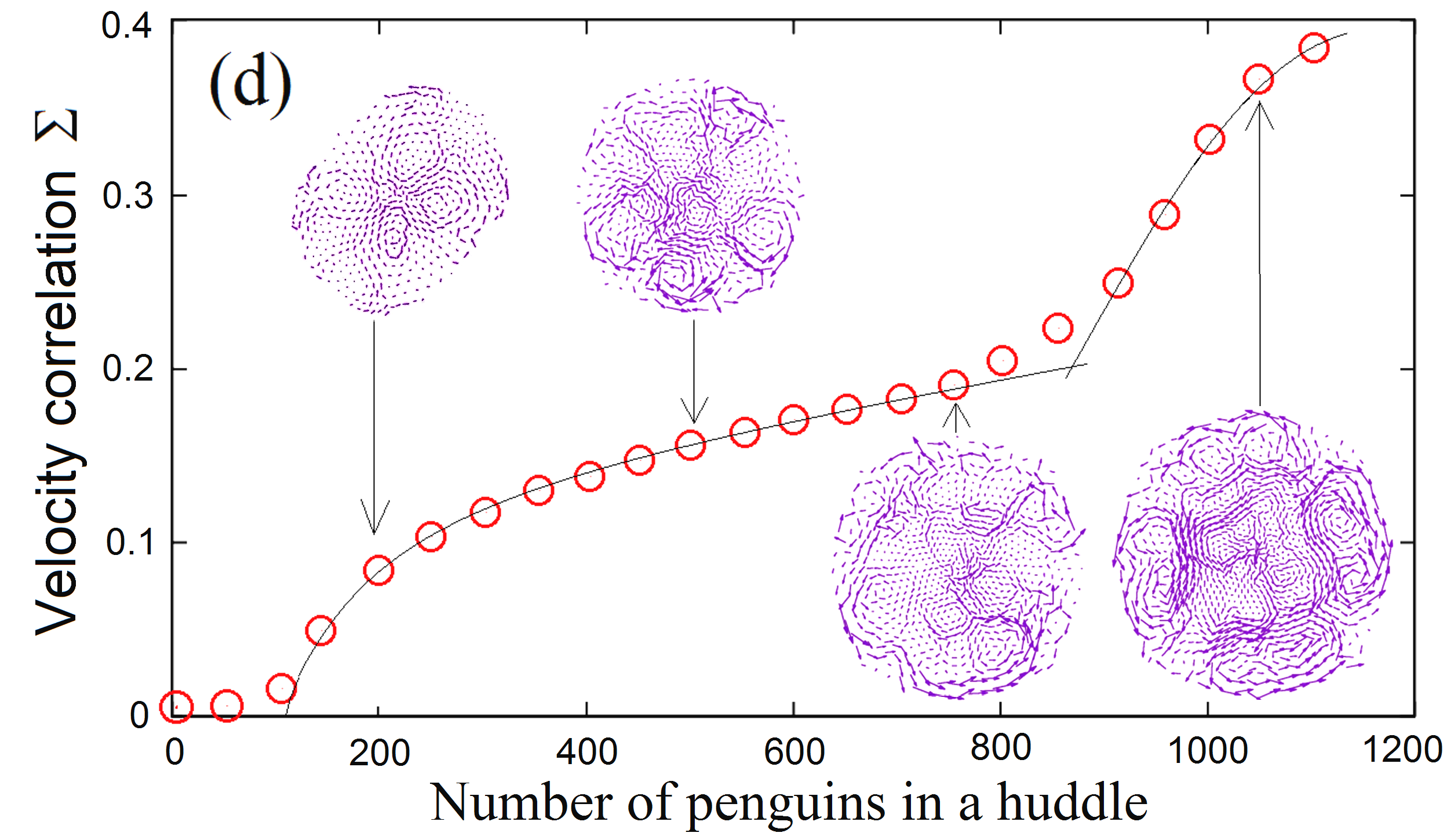}
\includegraphics[scale=0.215] {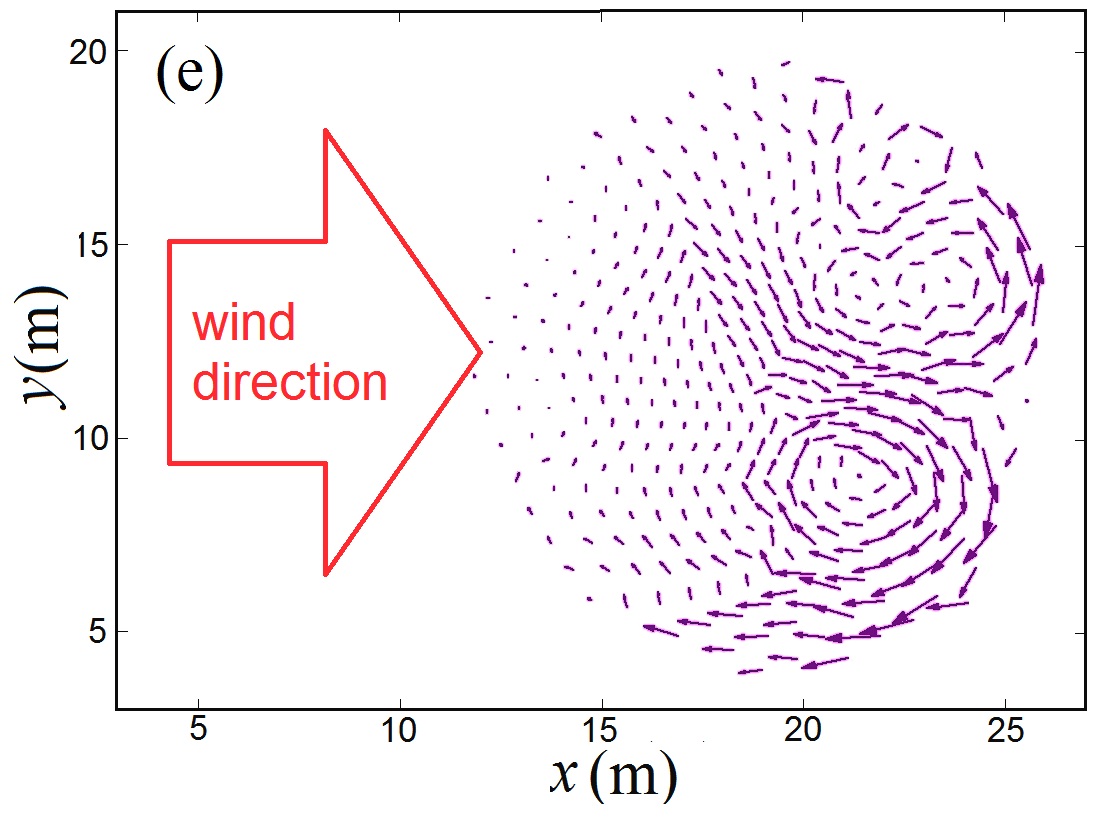}
\caption{\label{fig:1}
(a) Typical frame taken from film~\cite{Marche2005} demonstrating that a motionless huddle of penguins can experience spontaneous fluidisation and excitation of global circulations of birds on the scale of the entire flock; (b) The key assumption of our mathematical model: the intention of a penguin to move in the direction of a thermal gradient generates an efficient buoyancy in a huddle; (c) Individual-based model reproduces the assembling of 500 penguins into a dense huddle, in the center of which the temperature gradually reaches a comfortable value. The figure shows four consecutive frames of the time dynamics, indicating the instantaneous location of birds and the temperature distribution; (d) Bifurcation diagram showing the correlation function of penguin velocities averaged over an ensemble and realisations as a function of the number of participating birds. The inserts show the instantaneous velocity fields for some huddles; (e) With any wind speed, a two-vortex circulation of penguins replaces a motionless huddle.
The figure shows the result of a direct numerical simulation for a huddle of 500 penguins.}
\end{figure*}

%
%
{\it Microscopic modeling}.~-- Figure 1a shows a frame taken from film~\cite{Marche2005} demonstrating that an occasional huddle formed by penguins can be heterogeneous. Some birds are motionless, forming a quasi-crystalline structure ({\it solid state}), while others make poorly coordinated movements around this huddle ({\it gas}). One can also distinguish a part of the flock whose behavior is similar to that of an unstable fluid. One can observe the macroscopic two-vortex flow developed as a result of the instability.

We suggest that penguin interactions are best described by Aristotelian dynamics since each bird, while laying an egg, moves slowly and carefully, and its speed does not exceed $1$-$2$ km/h. Since only males of similar age participate in this process, we assume that all penguins have the same mass, denoted as $m$, and size, denoted as $R$. The system of equations describing the motions for $N$ penguins in the 2-D Cartesian coordinates is as follows:
\begin{eqnarray}
m{\dot{\bf r}}_i = k\nabla T\bigg|_{{\bf r}={\bf r}_i}+\sum_{j=1, j\ne i}^N e^\frac{2R-d_{ij}}{B} ,\quad {\bf r}_i (0) = {\bf r}_{0i} ,
\label{eq:1}
\\
\partial_t T = \chi \nabla^2 T - \lambda T - {\bf V}\cdot\nabla T + \sum_{j=1}^N Q_i ,\quad
T\bigg|_{{\bf r}\to\infty} = 0 ,
\label{eq:2}
\end{eqnarray}
where ${\bf r}_i$:($x_i$,$y_i$) is the position vector of the $i$-th penguin, $T$ is the ambient temperature, ${\bf n}_{ij}$ is the unit vector directed from the $j$-th to the $i$-th penguin, $d_{ij}$ is the distance between the mass centers of two birds; ${\bf V}$~is the wind speed, $\chi$ is the air thermal diffusivity, $\lambda$ is the vertical heat transfer coefficient, $Q$ is the rate of heat generation by penguins. $k$, $A$, and $B$ are the tuning parameters. The first term on the right-hand side of~(\ref{eq:1}) expresses the intention of the penguin to move in the direction of the heat gradient (Fig.1b). 
Although the emperor penguin is well adapted to low temperatures, this adaptation is insufficient during winter, resulting in the huddling phenomenon.
In essence, the force sets penguin thermotaxis, whose mechanism is similar to the phenomenon of chemotaxis in bacteria. However, even during huddling, penguins prefer to maintain some distance from each other because each bird is responsible for preserving the fragile egg. So, the second term in~(\ref{eq:1}) introduces a mechanism of weak repulsion between penguins~\cite{Helbing2000}. Eq.~(\ref{eq:2}) describes the dynamics of the flock thermal field, having standard terms of the heat conduction equation. In general, the Cauchy problem~(\ref{eq:1}) and the boundary value problem~(\ref{eq:2}) represent a closed system of coupled evolutionary equations that determine the dynamics of the thermal field and heat-generating agents in the collective thermotaxis. For more details, see {\it Supplementary materials}.

\begin{figure*}
\includegraphics[scale=0.10] {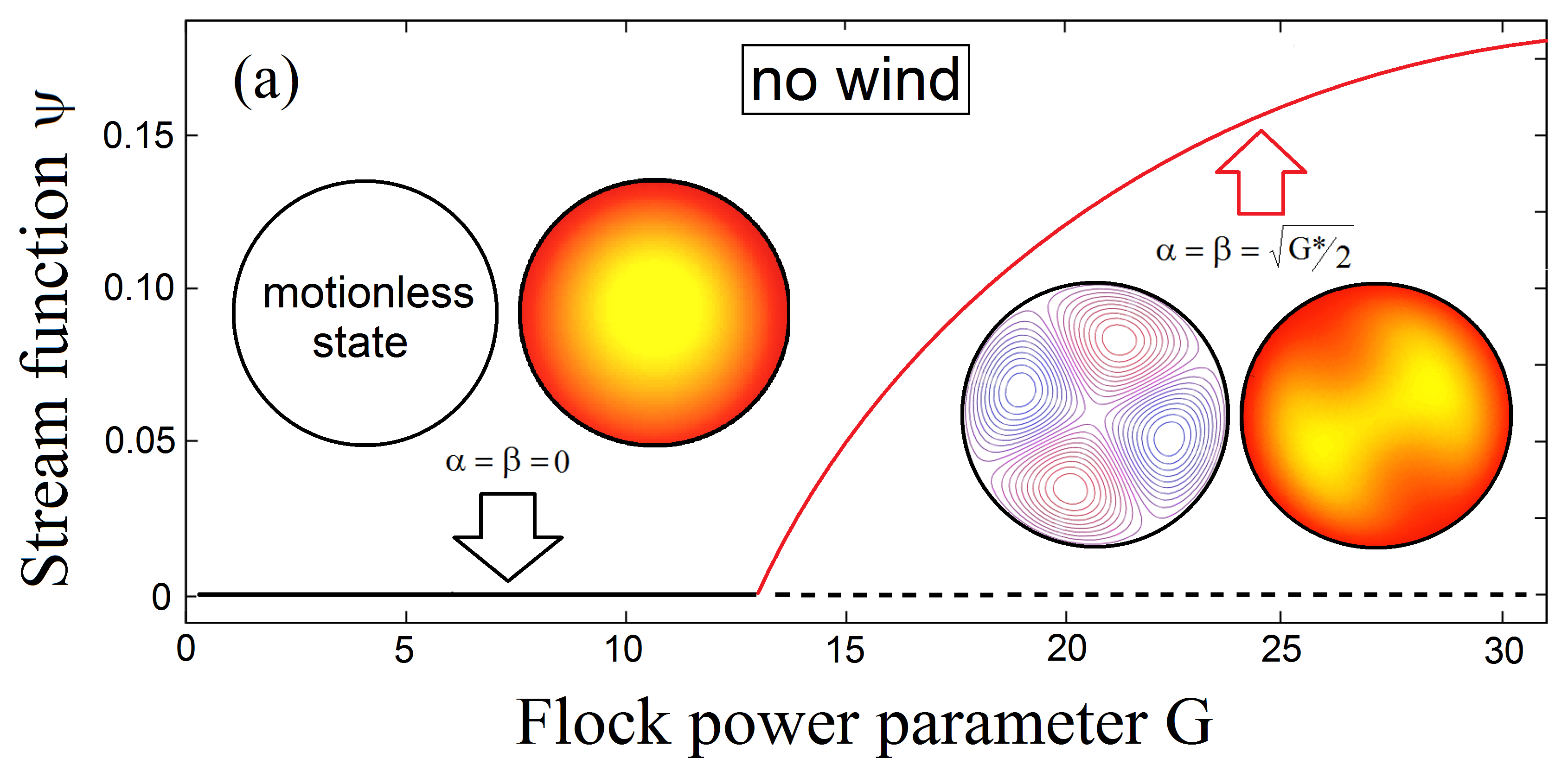}\hspace{7mm}
\includegraphics[scale=0.15]{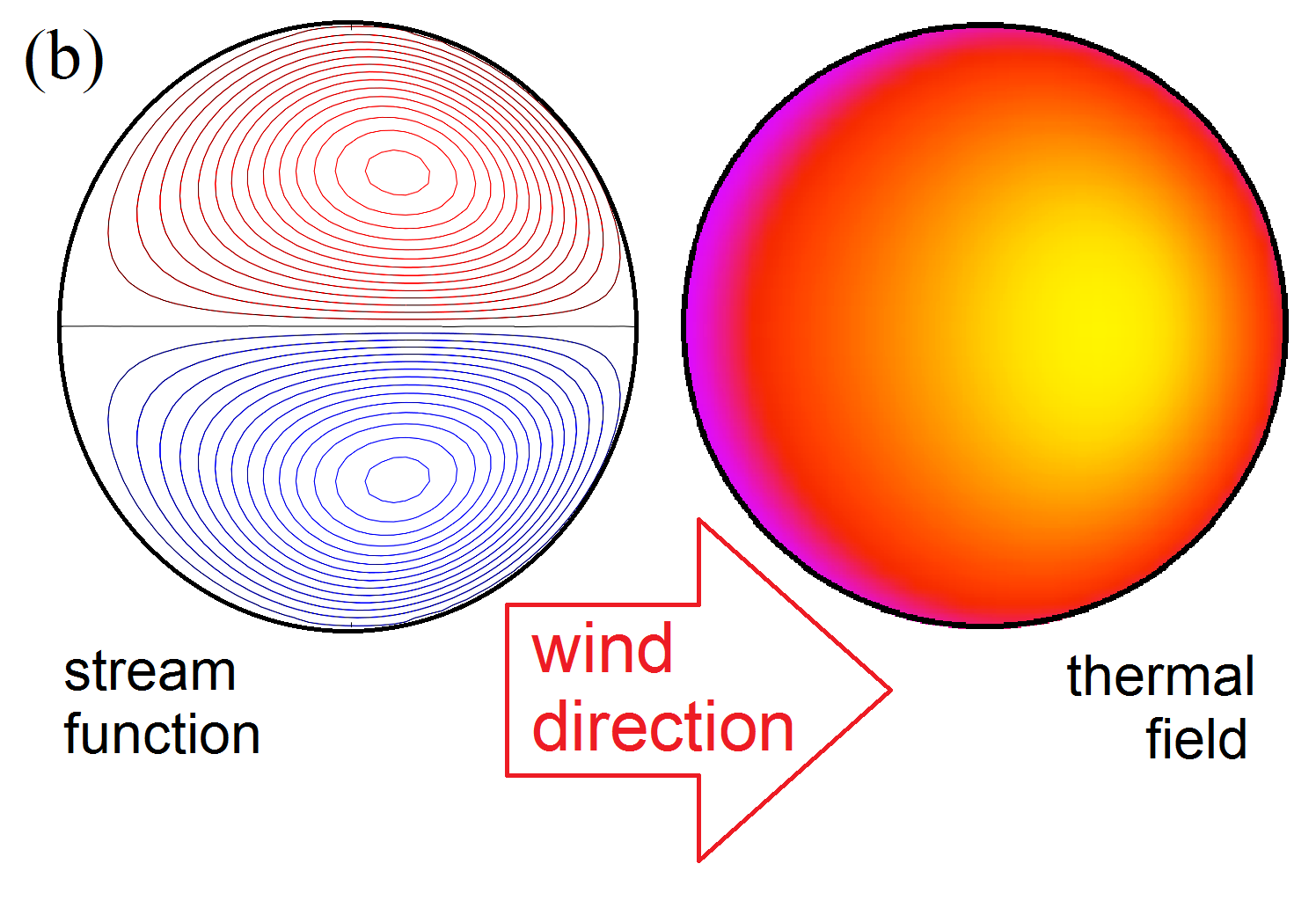}
\caption{\label{fig:2}
(a) Bifurcation diagram showing the branching of a steady-state solution which describes a four-vortex circulation of penguins in a round huddle with a gradual increase in the flock power parameter $G$. The diagram was obtained applying a weakly nonlinear analysis (see {\it Supplementary materials});
(b) Typical solution for the stream function (left) and huddle temperature (right) obtained by direct numerical simulation of problem~(\ref{eq:1}) for $G=14.3$, taking into account wind $W=0.1$.}
\end{figure*}

Fig.~1c shows the process of penguins huddling obtained in a numerical simulation of~(\ref{eq:1},\ref{eq:2}) for 500 penguins starting from random initial conditions. As in nature, the number of dense groups into which the birds unite may differ from one. The result depends on the initial distribution of the birds over the territory. One can notice that the larger the huddle, the higher the temperature in its center. Therefore, smaller groups finally break up. In a large huddle, the temperature in its center gradually reaches a comfortable value. 
Nonetheless, the thermal energy remains unevenly distributed, causing penguins at the edge of the still crowd to suffer from hypothermia.
The latter naturally pushes them to move closer to the huddle center. Our analysis of model (\ref{eq:1},\ref{eq:2}) revealed that the sudden fluidization of the penguin huddle results from the development of instability and occurs in a threshold manner. Fig.1d shows that the number of penguins participating in huddling is the bifurcation parameter of the problem.
There is a critical number of birds, about $100$ penguins, below which the huddle remains motionless. Bifurcation occurs when the crowd gets bigger. However, the huddle shape also influences the development of instability. If the primary crowd is significantly asymmetric and elongated, the onset of instability can be delayed. The first mode to lose stability is the four-vortex motion, as shown in Fig. 1d. 
Circulations of birds ensure a more uniform redistribution of thermal energy in the flock, as each penguin periodically moves from the edge of the flock to its center and back. We also found that, with an increase in the number of penguins in the liquefied part of the huddle, the circulation becomes smaller-scale and less ordered. It is interesting to note that the appearance of wind immediately transforms any motion into a two-vortex circulation (Fig.1e), while the convection threshold sharply drops. These results correlate well with natural observations (Fig.1a).

%
%
{\it Continuous modeling}. -- To verify that the spontaneous circulation of penguins is the result of convective instability, we developed a model of penguin behavior in a perfectly round huddle of radius $R_0$ under the continuous approximation. So, we consider a tightly packed crowd as an incompressible fluid moving in a highly dissipative medium under the action of the Darcy drag force~\cite{Bratsun1995}:
\begin{eqnarray}
\nabla\cdot{\bf u} = 0, \label{eq:3}\\
{\bf u} = -\nabla p + GT {\bf r}, \label{eq:4}\\
\partial_t T + ({\bf u}\cdot\nabla) T= \nabla^2 T - W ({\bf m}\cdot\nabla) T + 2G , \label{eq:5}\\
r=1:\quad {\bf u}\cdot{\bf r} = 0,\quad T = 0,\label{eq:6}
\end{eqnarray}
where $\bf u$ is the velocity, $p$ is the pressure.
Eqs.~(\ref{eq:3}-\ref{eq:6}) are non-dimensionalized using the units $R_0$, $R_0^2/\chi$, $\chi/R_0$, $\nu\chi\rho_0/K$, and $(\nu Q/2\rho_0 c_p g K)^{1/2}$ for length, time, velocity, pressure, and temperature, respectively. Two parameters
\begin{eqnarray}
G = R_0^2\sqrt{{qKQ}/{2\rho_0 c_p \nu \chi^2}},\quad W = {R_0 V}/{\chi} \label{eq:7}
\end{eqnarray}
are the huddle power and the wind speed, respectively.

In~(\ref{eq:4}), we introduce the centripetal force
\begin{eqnarray}
{\bf F} = -g(T(0)-T(r)) {\bf r} , \label{eq:8}
\end{eqnarray}
where $g$ is a tuning coefficient. 
It reflects the penguins' intention to approach the maximum of the thermal field. The force~(\ref{eq:8}) results in the appearance of the Boussinesq term in the motion equation~(\ref{eq:4}) and plays a crucial role in generating instability. In case of no wind and the motionless huddle, one can readily obtain an axisymmetric steady-state solution to problem~(\ref{eq:3}-\ref{eq:6}):
\begin{eqnarray}
T_0 (r) = G(1-r^2)/2   \label{eq:9}
\end{eqnarray}
shown in Fig.2a. We investigated the stability of~(\ref{eq:9}) using linear and weakly nonlinear analysis, as well as direct numerical simulation (see {\it Supplementary materials}). Fig.2a shows that if $W=0$, the excitation of penguins' circulation occurs by varying the flock power parameter $G$. As in microscopic modeling, the first critical mode is a four-vortex circulation (compare Figures 1d and 2a). As $G$ increases, the system dynamics quickly become non-stationary, including the transition to deterministic chaos.
In the case of wind, the motionless state~(\ref{eq:9}) does not exist since the axisymmetric solution does not pass through problem~(\ref{eq:3}-\ref{eq:6}). Therefore, $W$ stands for an imperfection parameter of the problem. Under even a weak wind effect $W\ne 0$, one observes a shift in the maximum thermal field and excitation of a two-vortex circulation (see Fig.2b). Notice that one can frequently see this type of penguins' motion in natural observations (Fig.1a).

To conclude, we demonstrated that emperor penguins survive the Antarctic winter through a physical mechanism of thermal convection.
To survive, penguins do not need to build any complex social relationships with each other. During a sudden cold snap, all they have to do is get together and form a tight crowd with enough strength. 
Then physics will activate the instability mechanism, leading to continuous circulations of birds in a flock.  
A similar phenomenon occurs during the bioconvection of bacteria in solutions. 
Although penguins are not bacteria, their collective behavior should not be surprising.
In extreme conditions, higher organisms can act reflexively, obeying simple but working physical mechanisms. Ultimately, this work demonstrates that self-organization in complex systems, which include simple elements like atoms and molecules, as well as more complex living beings such as bacteria and higher animals, perhaps even humans, operates under similar principles.

The study was supported by the Russian Ministry of Science and Higher Education (No. FSNM-2025-0001).

\end{document}